\documentclass[a4paper,twoside]{jpconf-mod}
\usepackage{graphicx}
\usepackage{cite}

\def\Lbabar{\mbox{{\LARGE\sl B}\hspace{-0.15em}{\Large\sl A}\hspace{-0.07em}{\LARGE\sl B}\hspace{-0.15em}{\Large\sl A\hspace{-0.02em}R}}}

\def\babar{\mbox{{\small \sl B}\hspace{-0.4em} {\small \sl A}\hspace{-0.03em}{\small \sl B}\hspace{-0.4em} {\small \sl A\hspace{-0.02em}R}}}

\newcommand\T{\rule{0pt}{2.6ex}}       
\newcommand\B{\rule[-1.2ex]{0pt}{0pt}} 

\def\ra{\rightarrow}

\def\gev{$\rm GeV$}
\def\gevc{$\rm GeV/c$}
\def\gevc2{$\rm GeV/c^2$}
\def\mev{$\rm MeV$}
\def\mevc{$\rm MeV/c$}
\def\mevc2{$\rm MeV/c^2$}
\def\fb{$\rm fb^{-1}$}

\begin{document}
\title{Direct Searches for New Physics Particles at \Lbabar}

\author{Gerald Eigen, \\ representing the \babar\ collaboration}

\address{Dept. of Physics, University of Bergen, Allegation 55, Bergen, Norway}

\ead{gerald.eigen@ift.uib.no}

\begin{abstract}
We present recent \babar\ results on searches for dark photons, long-lived  scalar particles and new $\pi^0$-like particles.
\end{abstract}

\section{Introduction}

\begin{figure}[h]
\centering
\vskip -0.9cm
\includegraphics[width=90mm]{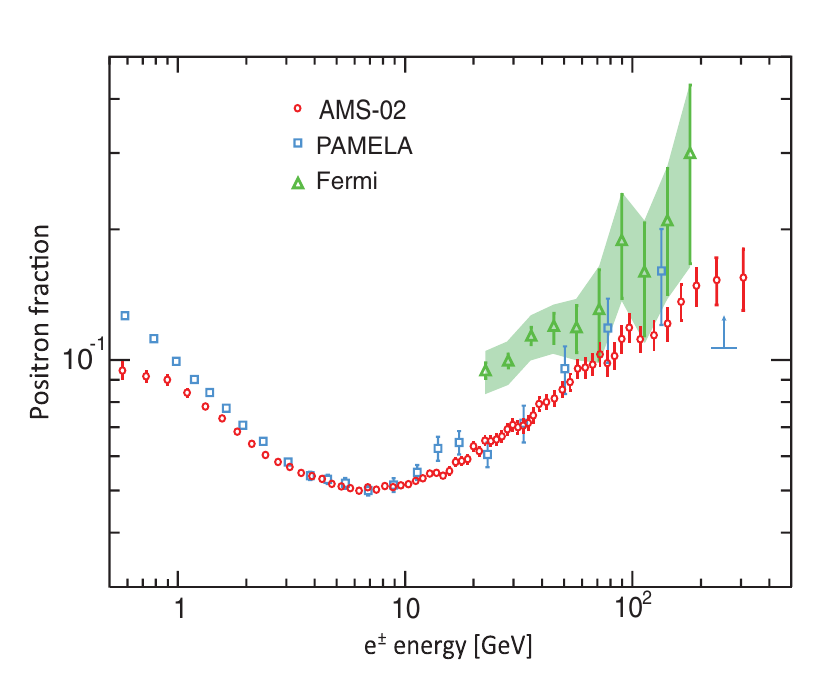}
\caption{Positron excess observed by Pamela, Fermi and AMS.}
\label{fig:positron}
\end{figure}

In 2010, the Pamela experiment reported a positron excess for energies greater 10~\gev\ that is increasing with energy~\cite{Pamela}. The excess
was confirmed by the Fermi Large Area Telescope  in 2012~\cite{Fermi}. Two years later, the AMS experiment published higher-precision data up to
300~\gev\ confirming an excess of the positron fraction in the 10 -- 250~\gev\ energy range~\cite{AMS}. 
Figure~\ref{fig:positron} shows the measurements from Pamela, Fermi and AMS. In the same energy range, Pamela and AMS did not observe any antiproton excess. Though an astrophysical explanation is very likely, theorists have come up with new dark-matter scenarios favoring light particles. In the past, dark-matter particles have been associated with new heavy particles predicted in extensions of the Standard Model (SM), such as neutralinos in supersymmetric models. 
Physicists have been searching for weakly-interacting massive particles (WIMPs) in dedicated experiments for many years~\cite{WIMP1, WIMP2, WIMP3, WIMP4, WIMP5, WIMP6,WIMP7, WIMP8, WIMP9, WIMP10, WIMP11, WIMP12, WIMP13, WIMP14, WIMP15, WIMP16, WIMP17}.
The observation of the positron excess has triggered searches for light dark matter in different accelerator experiments.

The SM may be connected to the dark sector through so-called portals~\cite{Essig}.  These are lowest-dimensional operators that may provide coupling of the dark sector to the SM\footnote{higher-dimensional operators are mass suppressed.}. Table~\ref{tab:coupling} lists these operators for couplings to new vector particles, pseudoscalars, scalars, and neutrinos.  At low-energy $e^+e^-$ collisions, the light vector portal is the most accessible portal, but the scalar portal can be probed as well~\cite{Pospelov}. In the simplest realizations of the dark sector, a new $U(1)^\prime$ symmetry is introduced~\cite{Holdom}. The associated gauge boson, called dark photon $(A^\prime)$, couples to the weak hyper charge with a mixing strength $\epsilon$. An effective interaction between the dark photon and the electromagnetic current arises after electroweak symmetry breaking. For an Abelian interaction, this mechanism would explain WIMP annihilation into SM fermions~\cite{Pospelov, Weiner, Arkani}. To accommodate the recent anomalies in cosmic rays, the dark photon mass is constrained to lie in the \mevc2\ to \gevc2\ range.

In recent years, we have searched for new physics in $e^+ e^-$ collisions collected with the \babar\ experiment at center-of-mass (CM) energies around 10~\gev~\cite{babar2, babar13c}.  
We searched for a dark Higgs $h^\prime$ plus a dark photon $A^\prime$ in the process $e^+ e^- \ra h^\prime A^\prime, h^\prime \ra A^\prime A^\prime$. The cross section is proportional to the product of fine structure constant $\alpha_D$ and $\epsilon^2$. Assuming a flat prior, we set $90\%$ confidence level (CL) Bayesian upper limits on $\alpha_D \cdot \epsilon^2$ of $10^{-10}$ to $10^{-8}$ 
for $h^\prime$ and $A^\prime$ masses in the  0.8 -- 10~\gevc2\ and 0.25 --  3~\gevc2\ regions, respectively~\cite{babar12a}. We also searched for dark non-Abelian gauge bosons in $ e^+ e^- \ra W^\prime W^\prime \ra e^+ e^- e^+ e^-, e^+ e^- \mu^+ \mu^-,  \mu^+ \mu^-  \mu^+ \mu^-$~\cite{babar09b}.

Furthermore, we  searched for  a new pseudoscalar $A^0$ in 
$\Upsilon(2S, 3S)$ radiative decays: 
$A^0 \ra \mu^+ \mu^-$~\cite{babar09c, babar13}, $\tau^+ \tau^-$~\cite{babar13a}, hadrons~\cite{babar11}, and invisible particles~\cite{babar11a}.  For the $\mu^+ \mu^-$ decay mode, we set $90\%$ CL Bayesian upper limits on the branching fraction of $0.26 \times 10^{-6}$ to $8.3 \times 10^{-6}$ for $A^0$ masses between 0.21~\gevc2\ and 10.1~\gevc2\ ~\cite{babar13}. For the other modes, we obtained similar results.  We also studied the decay $\Upsilon(1S) \ra \gamma A^0, ~A^0 \ra gg, ~s \bar s$ setting $90\%$ CL upper limits on the branching fraction at a level of $10^{-6}$ to $10^{-2}$ for $A^0$ masses in the 1.5 -- 9.0~\gevc2\ range \cite{babar13b}.

\begin{center}
\begin{table}[h]
\centering
\caption{\label{tab:coupling} Portals for connecting the dark sector to the Standard Model where $F^Y_{\mu \nu}$ is the strength of the Yang-Mills field, $F^{\prime \mu \nu}$ is the strength of the dark photon field, $\epsilon$ is the mixing parameter of a dark photon to a real photon, $a$ is the axion field, $f_a$ is the scale at which the Peccei-Quinn global $U(1)$ symmetry is spontaneously broken~\cite{Peccei, Weinberg, Wilczek}, $G^{\mu \nu}$ is the gluon field strength, $S$ is a dark scalar field with coupling strengths ($\lambda$ and $ \mu$) to the Higgs field $H$, and $N$ is the sterile neutrino field with coupling $y_N$ to the Higgs field.  } 
\vskip 0.2cm
\begin{tabular}{|l|c|c|}
\hline
Portal & Particle  & Operator \T \B \\
\hline
\hline
"Vector" &  Dark photons& $\epsilon  F^Y_{\mu \nu} F'^{\mu \nu}$   \T \B \\
"Axion" & Pseudoscalars & $\frac{a}{f_a} F_{\mu \nu} \tilde F^{\mu \nu}, \frac{a}{f_a} G_{i \mu \nu} \tilde G_i^{\mu \nu},  \frac{\partial_\mu a}{f_a} \bar \psi \gamma^\mu \gamma^5 \psi $  \T \B \\
"Higgs" &Dark scalars & $ (\mu S+ \lambda S^2) H^\dagger H$  \T \B \\
"Neutrino" &  Sterile neutrino & $y_N LHN$ \T \B\\
\hline
\end{tabular}
\end{table}
\end{center}

\section{Search for Dark Photons}

\begin{figure}[h]
\begin{minipage}{14pc}
\includegraphics[width=60mm]{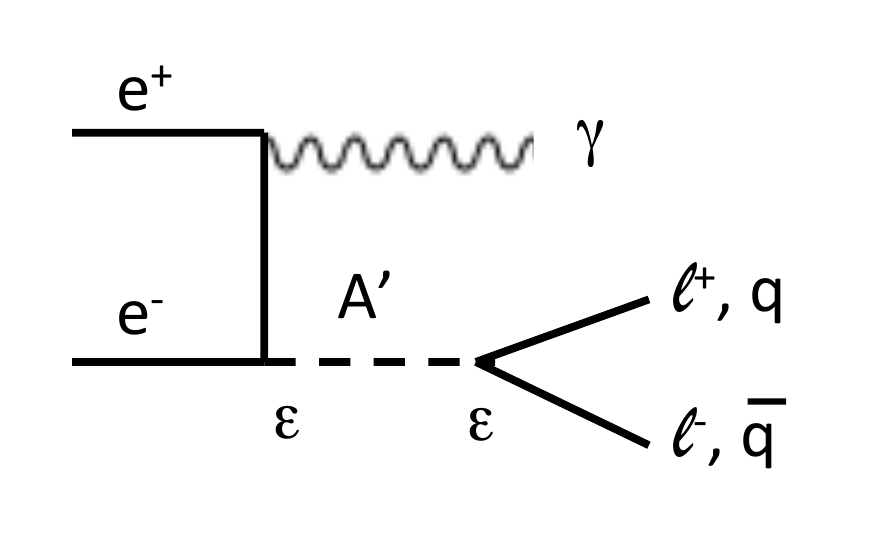}
\caption{\label{fig:gap}Lowest order diagram for $ e^+ e^- \ra \gamma A^\prime \ra  \gamma \ell^+ \ell^-, \gamma q \bar q$.}
\end{minipage}\hspace{2pc}%
\begin{minipage}{22pc}
\includegraphics[width=80mm]{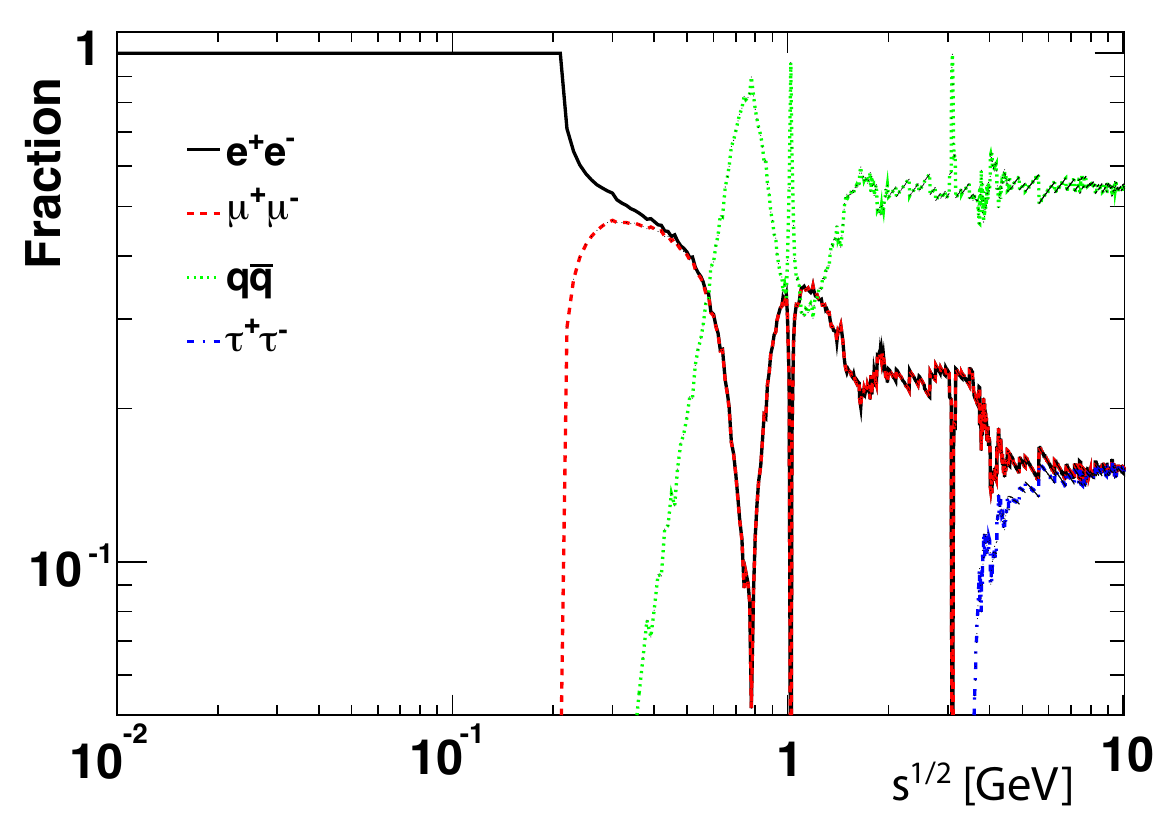}
\caption{\label{fig:dpdk}Branching fraction predictions for $A^\prime \ra \ell^+ \ell^-$ and $A^\prime \ra q \bar q$ as a function of $m_{A^\prime}$.}
\end{minipage} 
\end{figure}

In the mass region below 10~\gevc2,  $e^+ e^-$ collisions provide an excellent laboratory for producing dark photons $A^\prime$~\cite{Essig09, Batell}.  The processes $ e^+ e^- \ra \gamma A^\prime \ra  \gamma \ell^+ \ell^-$ ($\ell^\pm =e^\pm$ or $\mu^\pm$) depicted in Fig.~\ref{fig:gap} are well suited since they have high sensitivity for $A^\prime$ production because of high selection efficiencies and large $A^\prime$ decay rates as Fig.~\ref{fig:dpdk} shows~\cite{Batell}. The cross section with respect to $e^+ e^- \ra \gamma \gamma$ is reduced by $\epsilon^2$, where the mixing parameter $\epsilon$ is expected to lie in the range of $10^{-5}$ to $10^{-2}$. 
 
We have studied $e^+ e^- \gamma$ and $\mu^+ \mu^- \gamma$ final states at CM energies larger than 200~\mev\ using an integrated luminosity of 514~\fb of $e^+ e^-$ collisions~\cite{babar14}. The data were recorded at the $\Upsilon(4S), \Upsilon(3S), \Upsilon(2S)$ peaks and 40~\mevc2\ below the $\Upsilon(4S)$ peak with the \babar\ detector~\cite{babar2, babar13c} at the PEP-II asymmetric storage ring at the SLAC National Laboratory. We select two oppositely-charged leptons plus a photon and apply a constrained fit to the beam energy and interaction point (IP). We use additional kinematic constraints to improve purity, require good quality on the photon, electron and muon and remove electron conversions. We simulate the background processes $e^+ e^- \ra e^+ e^- (\gamma)$ and $e^+ e^- \ra \gamma \gamma (\gamma)$ with the generator BHWIDE~\cite{BHWIDE}, $e^+ e^- \ra \mu^+ \mu^- (\gamma)$ with the generator KK2F~\cite{KK2F} and resonance production in initial state radiation $e^+ e^- \ra \gamma X~ (X=J/\psi,~\psi(2S),~\Upsilon(1S),~\Upsilon(2S))$ using structure function techniques~\cite{Arbuzov, Caffo}.  For the simulation of the detector response, we use GEANT4~\cite{GEANT4}.

\begin{figure}[h]
\centering
\vskip -0.1cm
\includegraphics[width=78mm]{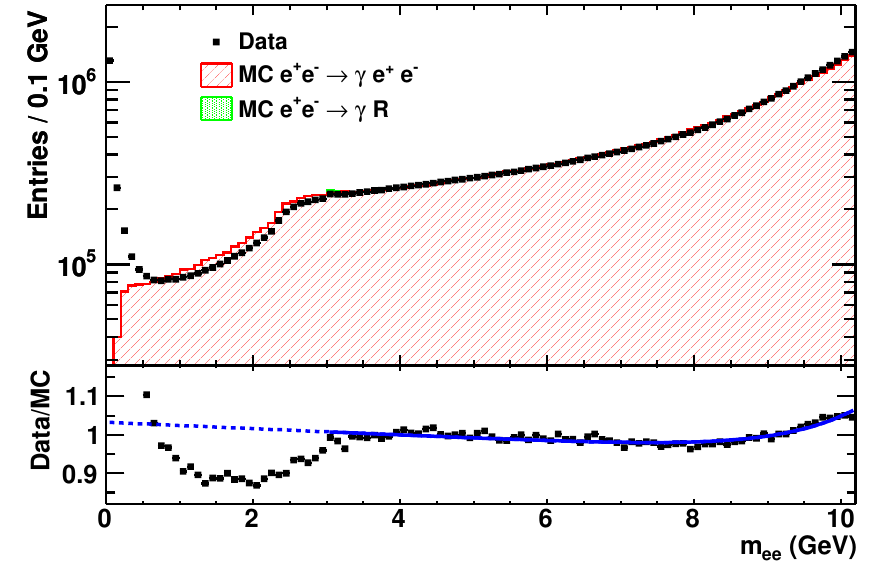}
\includegraphics[width=80mm]{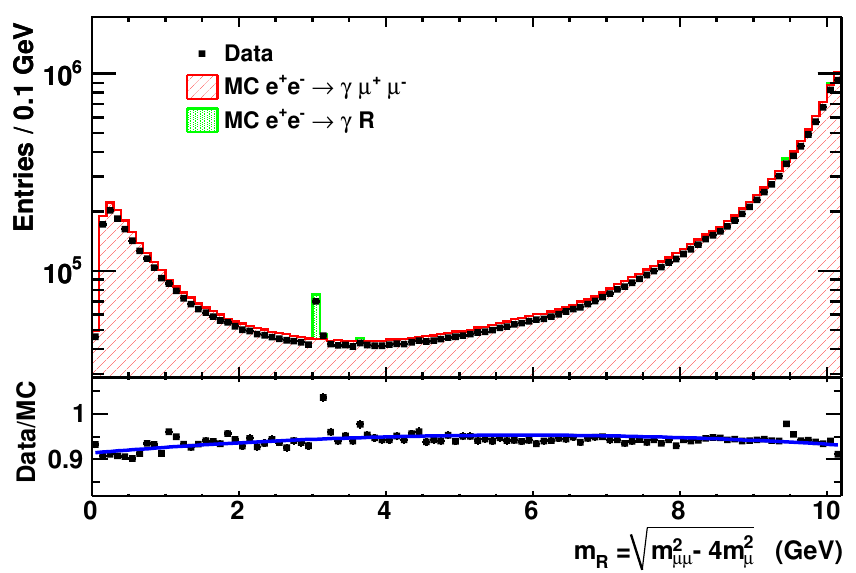}
\caption{Top: measured $m_{ee}$ (left) and $m_{R}$ (right) invariant-mass spectra in comparison to simulations; bottom left and right: the data/MC ratio.}
\label{fig:ll-mass}
\end{figure}

Figure~\ref{fig:ll-mass} (left) shows the dielectron invariant-mass spectrum in data and in simulation with the generator BHWIDE. Generally, the agreement between data and Monte Carlo (MC) simulation is good except for the low-mass region where the BHWIDE generator fails to reproduce the $ m_{ee}$ invariant-mass spectrum. On the other hand, the MADGRAPH generator~\cite{MADGRAPH} reproduces the low $m_{ee}$ region  in a consistent way but the limited sample size introduces too large uncertainties. Therefore, our signal extraction does not rely on the MC simulation.

For the dimuon sample, we define the reduced mass $m_{R}= \sqrt{(m^2_{\mu \mu} - 4 m^2_\mu)}$ 
 since the turn-on near threshold is smoother.
Figure~\ref{fig:ll-mass}  (right) shows the $m_R$ distribution in data and simulations. The KK2F generator~\cite{KK2F} reproduces $ m_{R}$  spectrum reasonably well. 
For both modes, we determine the selection efficiency from simulation yielding $\epsilon_{ee\gamma} =15\% $ and $\epsilon_{\mu \mu \gamma }=35\% $. The efficiency reduction in the $ee\gamma$ mode is due to mainly pre-scaling of radiative  Bhabha events in the trigger. However,  radiative Bhabhas are still the dominant background. Thus, the sensitivity for the dark photon search is dominated by the $\mu^+ \mu^-$ mode.
We extract the signal yield as a function of $m_{A^\prime}$ by performing independent fits to the $m_{ee}$ and $m_R$ mass distributions for each beam energy covering the range $0.02 < m_{A^\prime} < 10.2$~\gevc2\  and $0.212 < m_{A^\prime} < 10.2$~\gevc2 for the $ee\gamma$ and  $\mu \mu \gamma$ mode, respectively. In these fits, we vary the step size to about half the value of the expected mass resolution $\sigma_{m_{A^\prime}}$ and cover a mass range that is at least $20 \sigma_{m_{A^\prime}}$. The signal resolution is estimated by fits to a Gaussian function for various $A^\prime$ mass values and is interpolated to all other masses. The typical steps sizes vary between 1.5~\mevc2\ and 8~\mevc2\ yielding 5704 (5370) mass hypotheses for the $ee\gamma~ (\mu \mu \gamma)$ channel.

\begin{figure}[h]
\centering
\vskip -0.4cm
\includegraphics[width=90mm]{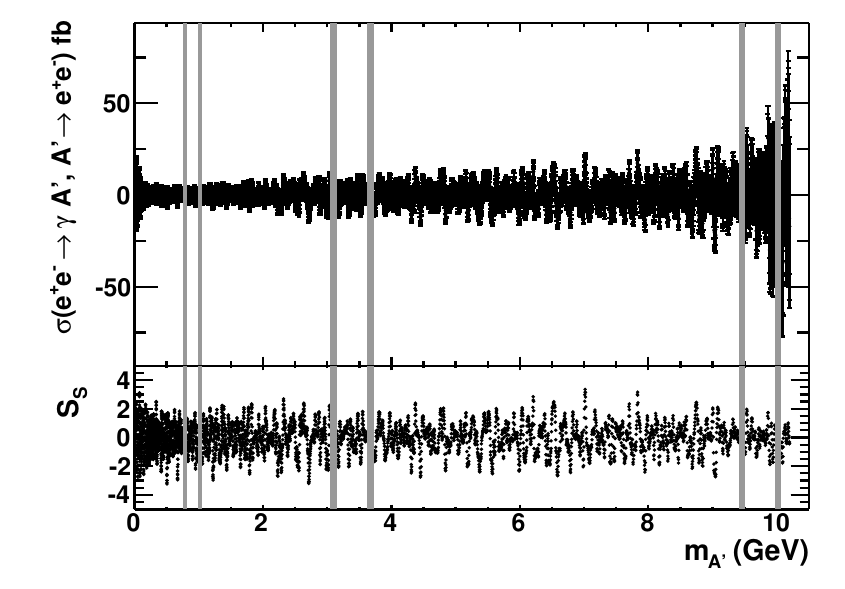}
\includegraphics[width=66mm]{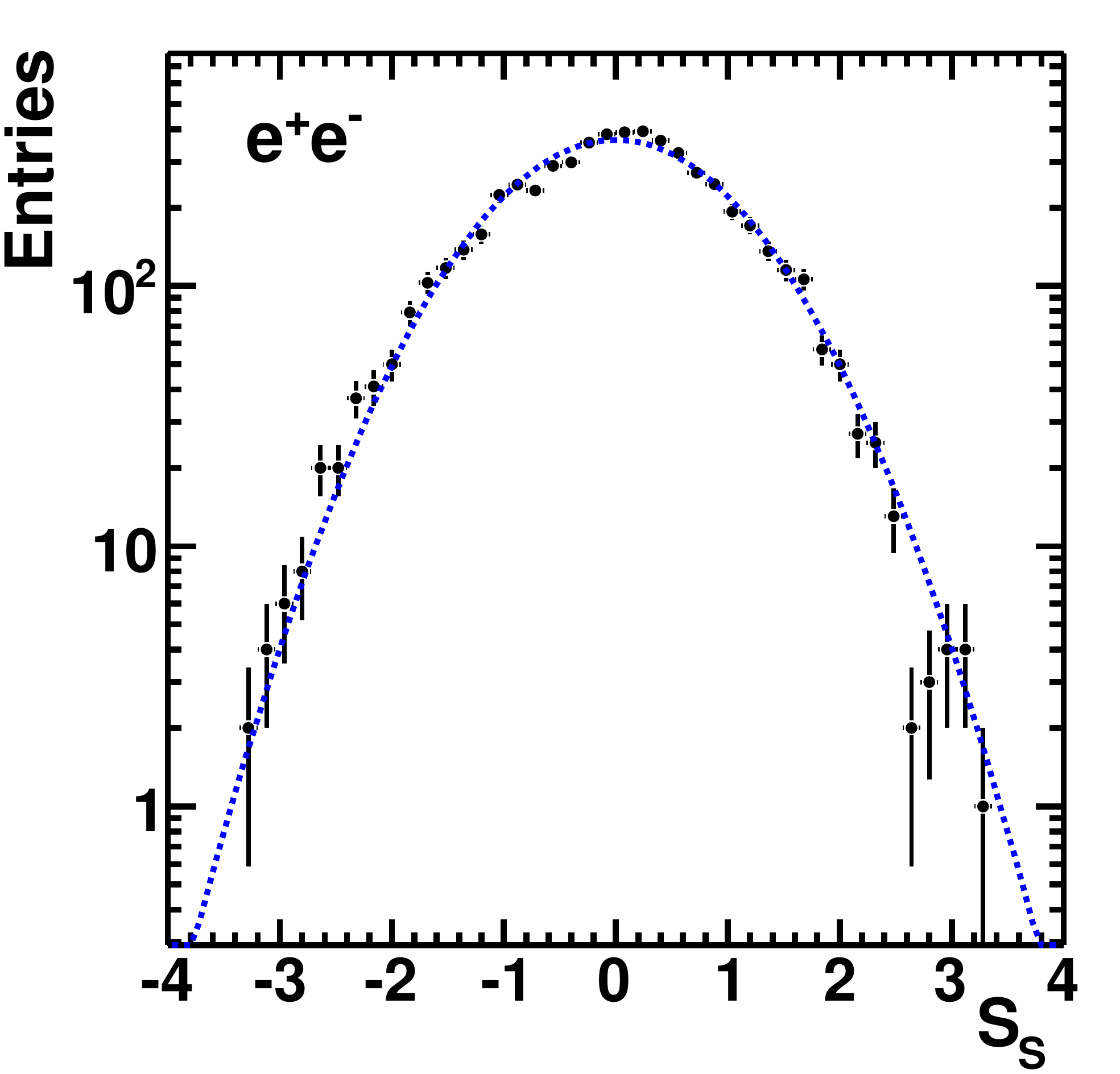}
\caption{Measurement of $\sigma(e^+ e^- \ra \gamma A^\prime \ra \gamma e^+ e^-)$ (left) and significance distribution (right). Grey vertical lines show the excluded regions around the vector mesons.}
 \label{fig:eeg}
\end{figure}

\begin{figure}[h]
\centering
\vskip -0.1cm
\includegraphics[width=90mm]{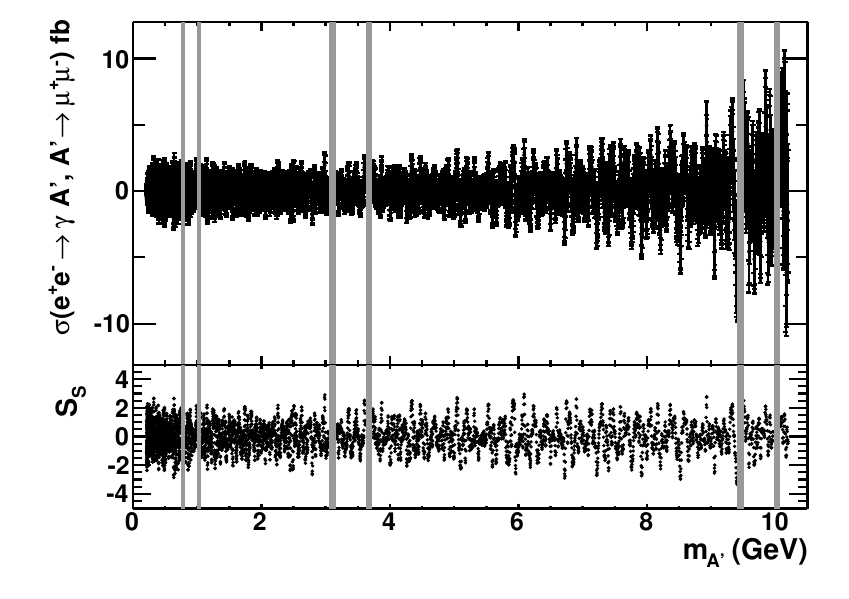}
\includegraphics[width=66mm]{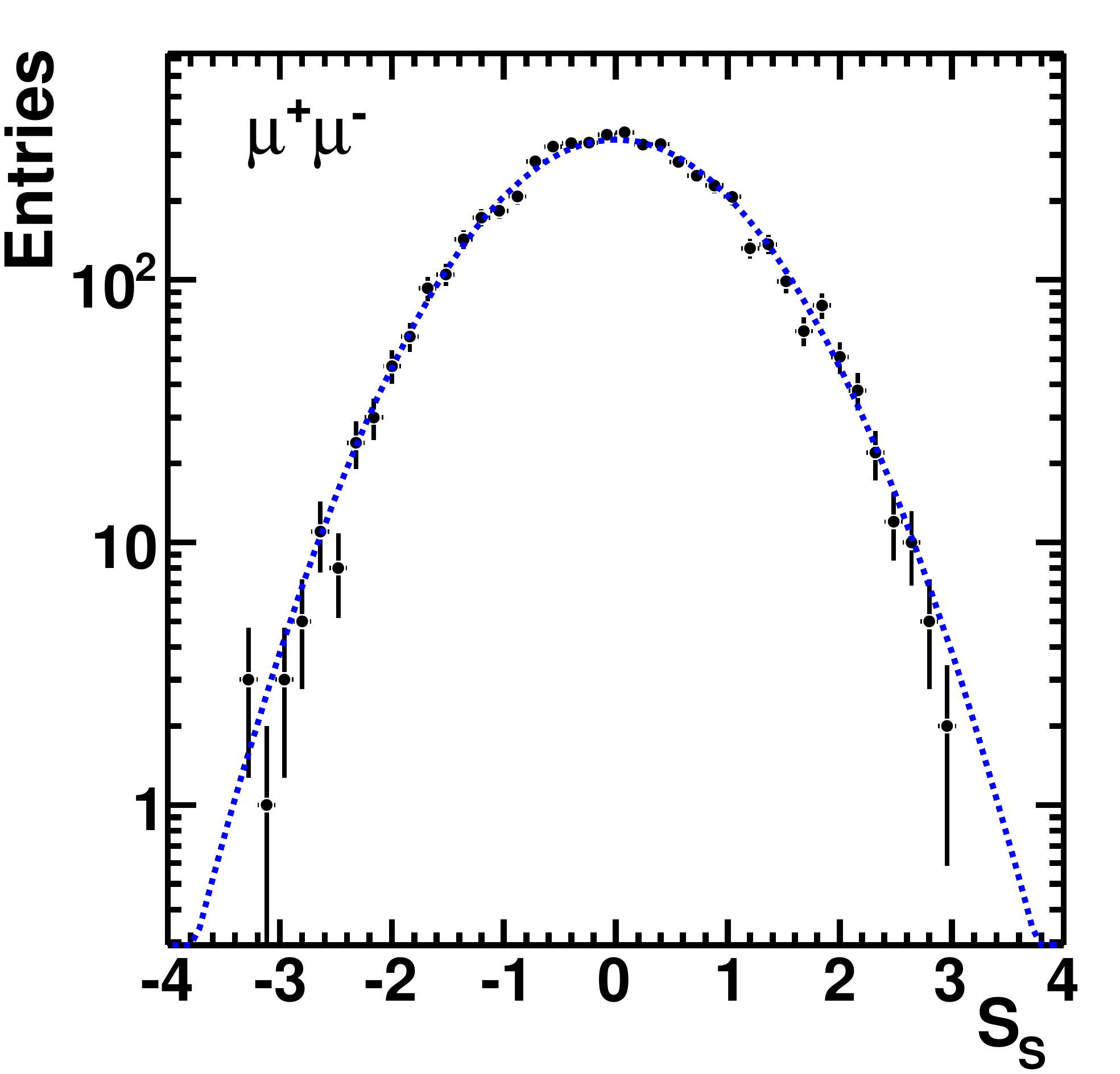}
\caption{Measurement of $\sigma(e^+ e^- \ra \gamma A^\prime \ra \gamma \mu^+ \mu^-)$ (left) and significance distribution (right). Grey vertical lines show the excluded regions around the vector mesons.}
 \label{fig:mmg}
\end{figure}

The likelihood function used to fit the observed spectra contains a signal component, radiative dilepton background and peaking background from vector mesons ($\omega,~ \phi,~ J/\psi,~ \psi(2S),~ \Upsilon(1S),~ \Upsilon(2S)$). The signal probability density function (PDF) is modeled directly from the simulated signal mass distribution by
a non-parametric kernel pdf and is interpolated between the known simulated masses using an algorithm based on the cumulative mass distributions~\cite{Read}. We estimate a systematic uncertainty of $5\%$ -- $10\%$ for this procedure.  The PDFs for the radiative dilepton backgrounds consist of higher-order polynomials. Though we parameterize the vector mesons $\omega, \phi, J/\psi, \psi(2S), \Upsilon(1S)$ and $\Upsilon(2S)$ with Gaussian and Crystal Ball functions, we exclude the mass regions around them in the dark photon search. We determine the $e^+ e^- \ra \gamma A^\prime \ra \gamma e^+ e^-$ and $e^+ e^- \ra \gamma A^\prime \ra \gamma \mu^+ \mu^-$ cross sections by dividing the extracted signal yield by efficiency, luminosity and dark photon decay branching fractions~\cite{Batell}. Figures~\ref{fig:eeg} (left) and \ref{fig:mmg} (left) show the measured cross sections as a function of $m_{A^\prime}$ for the $e^+ e^- \gamma$ and $\mu^+ \mu^- \gamma$ modes, respectively.  The results include all data recorded at the $\Upsilon(2S), \Upsilon(3S)$ and $\Upsilon(4S)$ except for $5\%$ data used to optimize the selection criteria. We estimate the significance for each fit by  ${\cal S}_S=\sqrt{2 {\cal L}/{\cal L}_0}$ where ${\cal L} $ is the likelihood for signal plus background while ${\cal L}_0$ is that for pure background. Figures~\ref{fig:eeg} (right) and \ref{fig:mmg} (right) show the resulting distributions for both modes based on 5704 and 5370 mass hypotheses, respectively. The largest deviation in the $e^+ e^-$ invariant-mass spectrum occurs at $m_{ee}=7.2$~\gevc2 with a significance of ${\cal S}_S =3.4 $. Including a trial factor determined from the MC, the significance is reduced to $0.6 \sigma$.  The largest deviation in the $\mu^+ \mu^-$ invariant-mass spectrum  occurs at $m_{\mu \mu}= 6.09$~\gevc2\ with a significance of  ${\cal S}_S=2.9 $. The application of a trial factor reduces this to $0.1\sigma$.

\begin{figure}[h]
\centering
\vskip -0.1cm
\includegraphics[width=110mm]{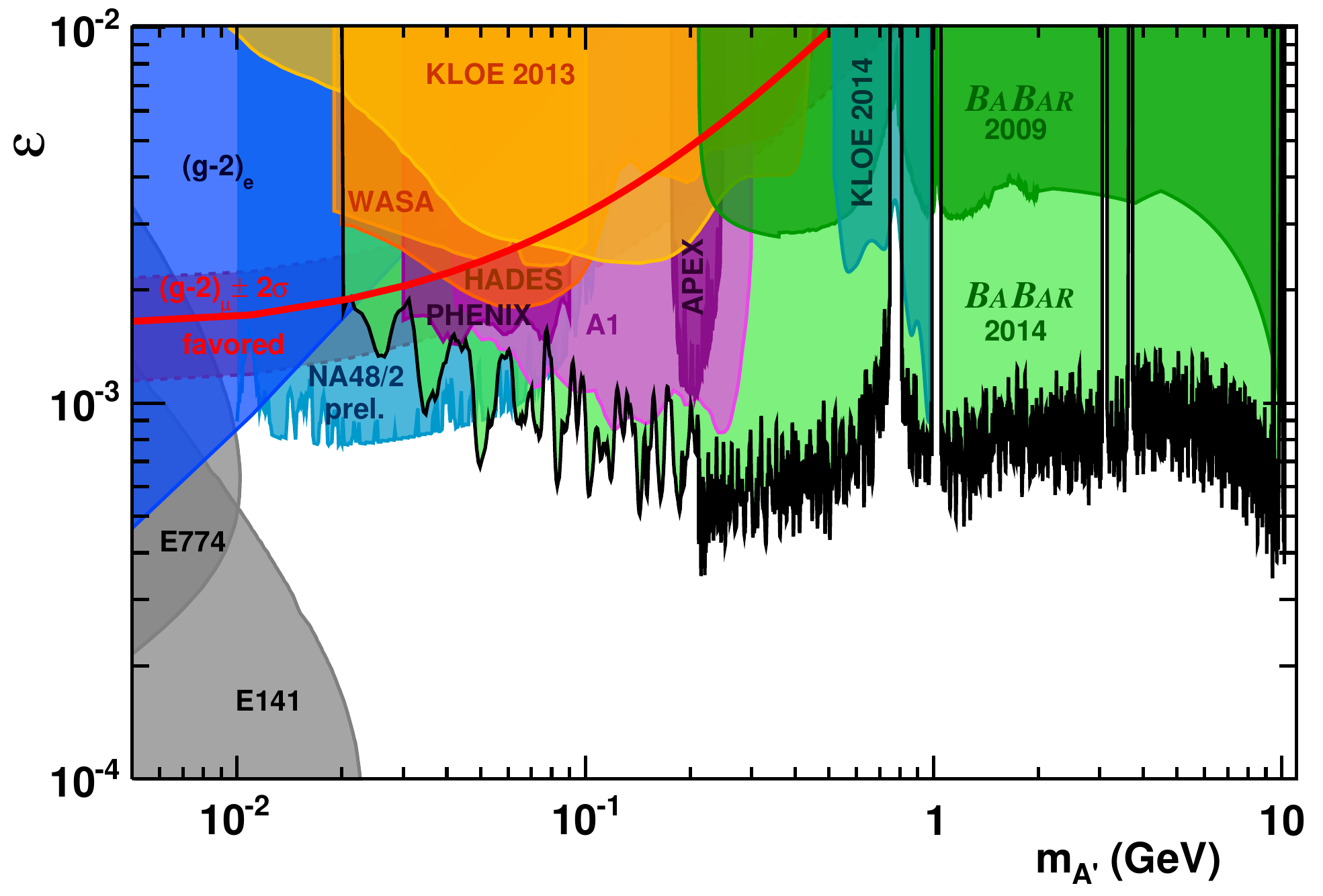}
\caption{Exclusion region of the mixing parameter $\epsilon$ as a function of the dark photon mass for present measurements.}
 \label{fig:llg-epsilon}
\end{figure}

\begin{figure}[h]
\centering
\vskip -0.1cm
\includegraphics[width=100mm]{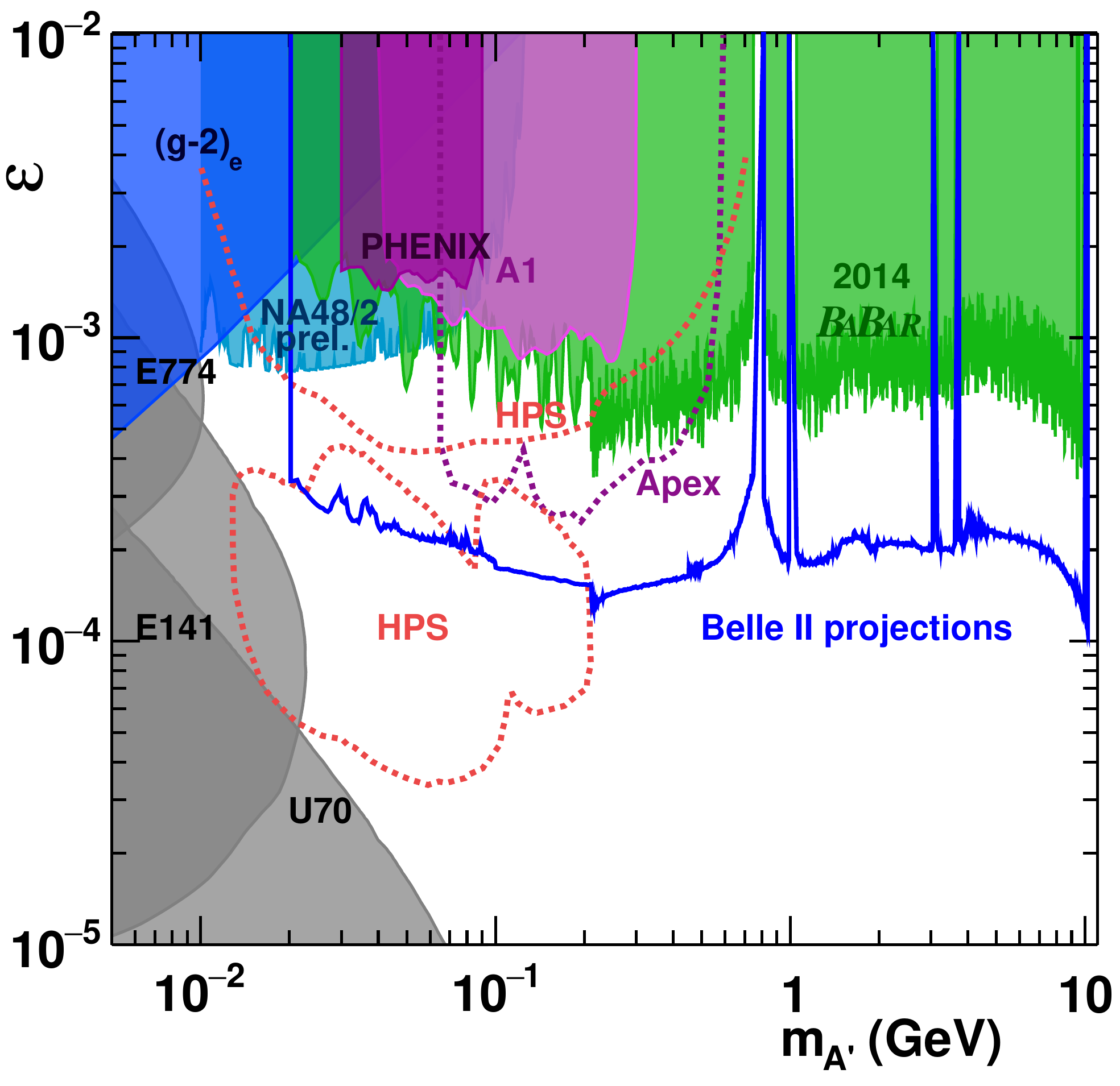}
\caption{Expected exclusion region on $\epsilon$ from results of new dedicated experiments APEX, DarkLight  and HPS, MESA, VEPP3 and Belle II.}
 \label{fig:llg-epsilon-fut}
\end{figure}

We combine the cross section measurements into a single result. Since we observe no signal, we determine Bayesian upper limits on the cross section at $90\%$ CL assuming a flat prior.  We include systematic uncertainties from the uncertainties of the dark photon branching fractions, luminosity and limited MC statistics. The resulting cross section upper limits are typically of the order of ${\cal O} (1$ -- 10~fb). We then translate these results into $90\%$ CL Bayesian upper limits on the mixing parameter as a function of the dark photon mass. 
Figure~\ref{fig:llg-epsilon} shows our results in comparison to dark photon searches of other experiments: KLOE~\cite{Babusci13, Babusci14} in $e^+ e^- \ra \mu^+ \mu^- \gamma$ and $\phi(1020) \ra \eta e^+ e^-$, electron-nucleon fixed target experiments APEX ~\cite{Abrahamyan} and A1~\cite{Merkel} in $A^\prime \ra e^+ e^-$, the proton fixed-target experiment HADES~\cite{Agakishiev} in $A^\prime \ra e^+ e^-$, WASA~\cite{Adlarson} and NA48~\cite{Goudzovski} in $\pi^0 \ra e^+ e^- \gamma$ and electron beam dump experiments~\cite{Andreas, Bross, Riordan}.
In addition, we show the constraints from the measurement of the anomalous electron magnetic moment~\cite{Endo}. The band of favored $\epsilon$ versus $m_{A^\prime}$ is obtained if the discrepancy  between the measurement and the SM calculation of the muon magnetic moment~\cite{Pospelov09} is attributed to dark photon production. In the 0.03 -- 10~\gevc2\ mass region, our results push the exclusion region  substantially lower and supersede the results of our previous analysis~\cite{babar09}.  Together with the results from NA48, we exclude the entire region favored by the dark-photon scenario for the g-2 measurements\footnote{This assumes that decays to invisible particles are small, otherwise the bounds are weakened.}.
By searching for $A^\prime \ra \pi^+ \pi^-$, we can further probe the region near the $\rho$ mass. 
Figure~\ref{fig:llg-epsilon-fut}  shows the exclusion region of $\epsilon$ as a function of $m_{A^\prime}$ that is expected from future measurements such as Belle II~\cite{belle2} for 50~$\rm ab^{-1}$ and  several new dedicated experiments including APEX, DarkLight  and HPS at Jefferson Laboratory~\cite{Boyce12, Balewski}, MESA at MAMI in Mainz~\cite{Molitor} and VEPP3 at Novosibirsk~\cite{Rachek}. Above 200~\mevc2, the Belle II data will reduce the exclusion limit on $\epsilon$ to $\sim 2 \times 10^{-4}$ while at lower masses HPS will push the limit below $10^{-4}$.

\section{Search for Long-Lived Particles}

The recent anomalous astrophysical observations have generated also interest in low-mass, long-lived hidden-sector states. 
Various models discuss this scenario, including those with dark photons~\cite{Batell, Essig09, Bossi, Schuster},  an inflaton~\cite{Bezrukov}, supersymmetry~\cite{Cheung, Schmidt}, a dark Higgs boson~\cite{Clarke} and with other dark-sector states~\cite{Nelson}. Several multipurpose experiments have conducted searches for long-lived particles in the sub~\gevc2~\cite{Andreas, Gninenko, Adams} and the multi-\gevc2 ~\cite{Abazov06, Abazov09, Abe, Aad13, Aad12, Aad13a} mass regions. In addition, dedicated experiments have been proposed~\cite{Essig11} and some are under construction~\cite{Moreno}. However, the ${\cal O}(\rm GeV/c^2)$ mass region has remained mostly unexplored.

\begin{figure}[h]
\centering
\vskip -0.4cm
\includegraphics[width=120mm]{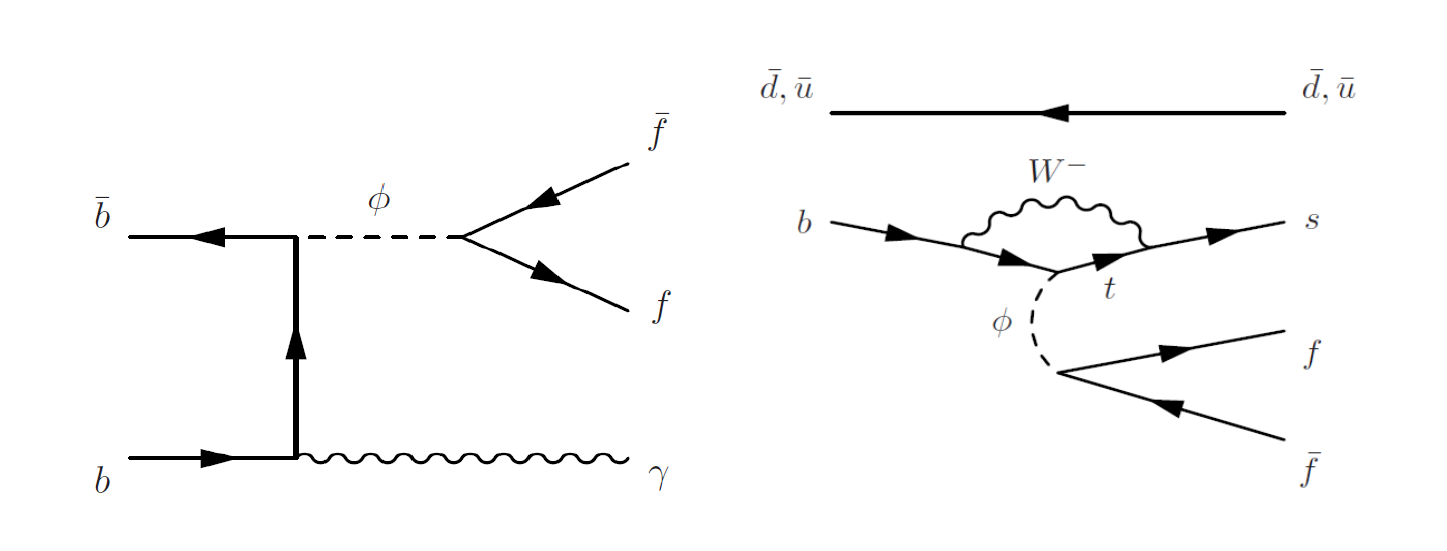}
\caption{Lowest-order diagrams for the production of a long-lived dark scalar particle in radiative $\Upsilon$ decays (left) and in penguin decays of $B$ mesons (right).}
 \label{fig:dark-scalar}
\end{figure}

At the $B$ factories, a hidden-sector scalar particle may be produced in an $\Upsilon$ radiative decay or in a $B$ penguin decay and, in turn, may decay into a pair of fermions as shown in Fig.~\ref{fig:dark-scalar}.  So we performed the 
first search for a long-lived particle L in the process $e^+ e^- \ra L X$ where X is any set of particles and L decays to six different final states. Except for 20~\fb taken at the $\Upsilon(4S)$, we use the entire \babar\ data collected at the $\Upsilon(4S)$, 40~\mev\ below the $ \Upsilon(4S)$ peak, at the $\Upsilon(2S)$,  and $\Upsilon(3S)$ corresponding to a luminosity of 489.1~\fb. We present the results in two ways. First without any assumptions on the production mechanism, we use the complete data sample and present model-independent results on the product of cross section, branching fraction and reconstruction efficiency of each two-body final state f, $\sigma_f=\sigma( e^+ e^- \ra L X) \cdot {\cal B}(L \ra f) \cdot \epsilon_f$. We have produced tables of the reconstruction efficiency as a function the L mass, transverse momentum $p_T$ in the CM frame and the proper decay distance $c \tau$ assuming L to be a scalar particle. 
Second, we present results for $B$ decays via $B \ra L X_s$ where $X_s$ is a hadronic state with strangeness S=-1. This production mechanism is motivated by the Higgs portal~\cite{Bezrukov, Cheung, Schmidt, Clarke} and axion-portal~\cite{Freytsis} models of dark matter. In this case, we report model-dependent results on the product branching fraction ${\cal B}_{Lf} = {\cal B}(B\ra L X_s) \cdot {\cal B}(L \ra f)$.

We use MC simulations of signal and background events to determine the signal efficiency and L mass resolution $\sigma_m$. We generate signal events with EvtGen~\cite{Lange} and produce two types of MC samples. In the model-independent approach, we generate $e^+ e^- \ra L n \pi$ and $B \ra L n \pi$ where $n \leq 3$. We generate 11 L mass values (0.5, 1, 2, 3, 4, 5, 6, 7, 8, 9, 9.5~\gevc2) assuming spin zero. This yields 
broad L-momentum spectra requiring efficiency tables to test specific models~\cite{babar15}. For the model-dependent approach, we produce $e^+ e^- \ra B \bar B, ~B \ra X_s L$ events where $X_s$ is composed of $10\% ~K$, $25\% ~ K^*$ and $65\% ~K^*(1680)$~\cite{pdg}. The high-mass tail of the $X_s$ mass spectrum is suppressed by phase space limitations of heavy L-states. This choice yields an L-momentum spectrum as a function of $m$ that reproduces the dimuon invariant-mass spectrum for $B \ra  X_s \mu^+ \mu^-$~\cite{Lange}. In this approach, we generate seven mass values (0.5, 1, 2, 3, 3.5, 4, 4.5~\gevc2). In addition, we generate the background processes $e^+ e^- \ra B \bar B$ (EvtGen~\cite{Lange}), $ \tau^+ \tau^-, \mu^+ \mu^-$ (KKF2~\cite{KK2F}), $e^+ e^-$ (BHWIDE~\cite{BHWIDE}) and $ q \bar q$ (JETSET~\cite{JETSET}) where $(q=u, d, s, c)$. The detector response is simulated with GEANT4~\cite{GEANT4}.

We reconstruct L from two oppositely-charged particles ($e^+ e^-, \mu^+ \mu^-, e^\pm \mu^\mp, \pi^+ \pi^-, K^+ K^-$ and $ K^\pm \pi^\mp$)
 that originate from a common vertex separated from the IP by more than $3 \sigma_{PV}$ where $\sigma_{PV}$ is primary vertex resolution.  We reject background from $K^0_S$ and $\Lambda$ decays as well as low-mass peaking structures  in the background by imposing mass thresholds:
$m_{ee} > 0.44$~\gevc2, $m_{e\mu} > 0.48$~\gevc2, $m_{\mu \mu} > 0.50$~\gevc2\ and $m_{\mu \mu} < 0.37$~\gevc2,   $m_{\pi\pi} >  0.86$~\gevc2, $m_{K\pi}> 1.05$~\gevc2\ and  $m_{KK} > 1.35$~\gevc2. 
The efficiency varies from $47\%$ for $m=1$~\gevc2, $2< p_T  <3$~\gevc2\ and $ c\tau = 3$~cm  to nearly zero for large $m$ and large $c \tau$ values.  

\begin{figure}[h]
\centering
\vskip -0.4cm
\includegraphics[width=110mm]{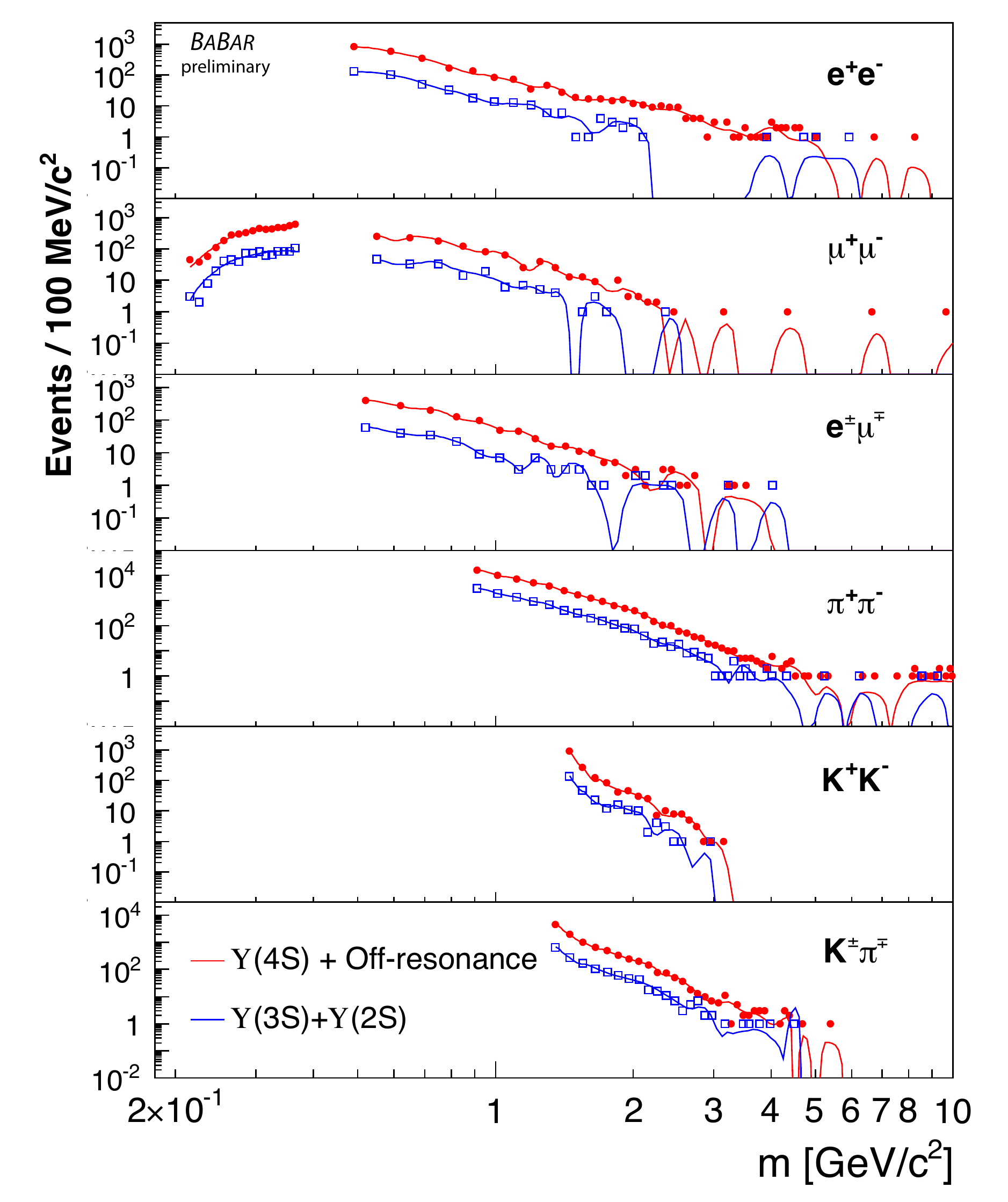}
\caption{Observed mass distribution for $\Upsilon(4S)$+ off-resonance data (red solid points),  for $\Upsilon(3S)+ \Upsilon(2S)$ data (blue open squares) as a function of mass $m$ for each final state with the background PDF $P_B$ overlaid (red, blue solid lines). }
 \label{fig:yields}
\end{figure}

We extract the signal yield for each final state as a function of $m$ using unbinned extended maximum likelihood fits of the $m$ distributions. The strategy is based on the fact that a signal would produce a peak in the $m$ distribution whereas background varies smoothly. Thus, the signal 
PDF is the pull histogram $P_S^i(m)=H^i_S((m_t -m)/\sigma_m)$ produced from masses closest to the true mass $m_t$, measured mass m and its uncertainty $\sigma_m$. The background PDF $P_B^i(m)$ is a second-order polynomial spline with knots located at the bin boundaries and is determined from the data. The bin width is variable and is defined as 15 times the RMS width of the signal distribution in that region. The factor 15 is chosen since it is sufficiently large to prevent $P^i_B(m)$ conforming to signal peaks but is sufficiently small to produce signal peaks from fluctuations. We scan the data in steps of 2~\mevc2\ using the PDF $n_S \cdot P_S(m) + n_B \cdot P_B(m)$ where $n_S~ (n_B)$ is the signal (background) yield determined from the fit. Figure~\ref{fig:yields} shows the observed number of events in comparison to the expected background for each final state. 
Two mass points in the dimuon final state show yields with significance ${\cal S}_S>3$. At $m_t =0.212$~\mevc2, we see 13 events with ${\cal S}_S=4.7$. However, most of the vertices lie inside the detector material and the muon tracks have low momenta that are poorly separated from e and $\pi$. At $m_t =1.24$~\gevc2, we see ten events with ${\cal S}_S=4.2$. We determine a probability of $P=0.04\% (0.8\%)$ of seeing yields with ${\cal S}_S \ge 4.7 ({\cal S}_S \ge 4.2)$ in the entire dimuon mass region. Including the other final states, the probabilities are reduced by a factor of six. Thus, the data are consistent with background expectations.

\begin{figure}[h]
\centering
\vskip -0.4cm
\includegraphics[width=79mm]{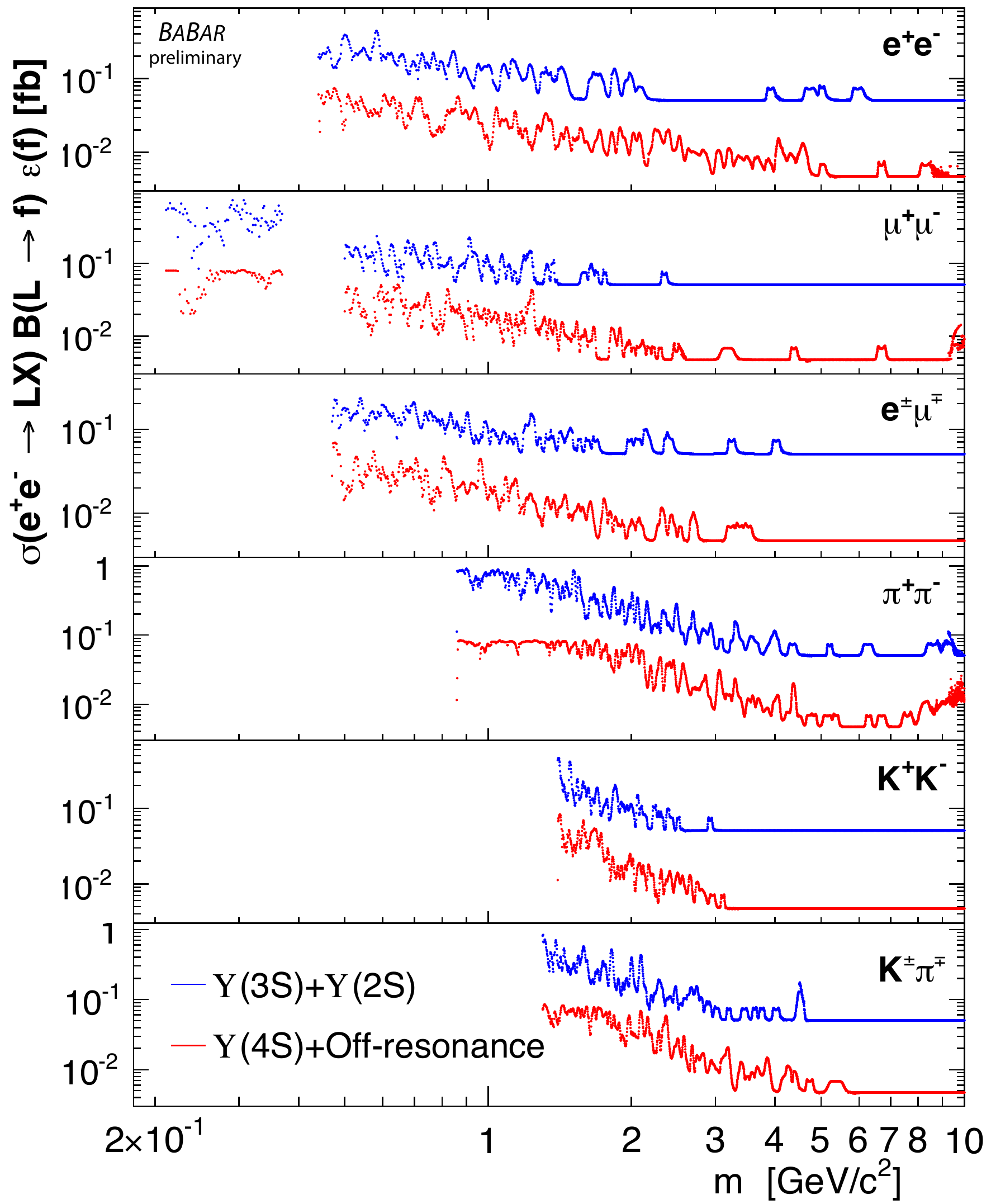}
\includegraphics[width=79mm]{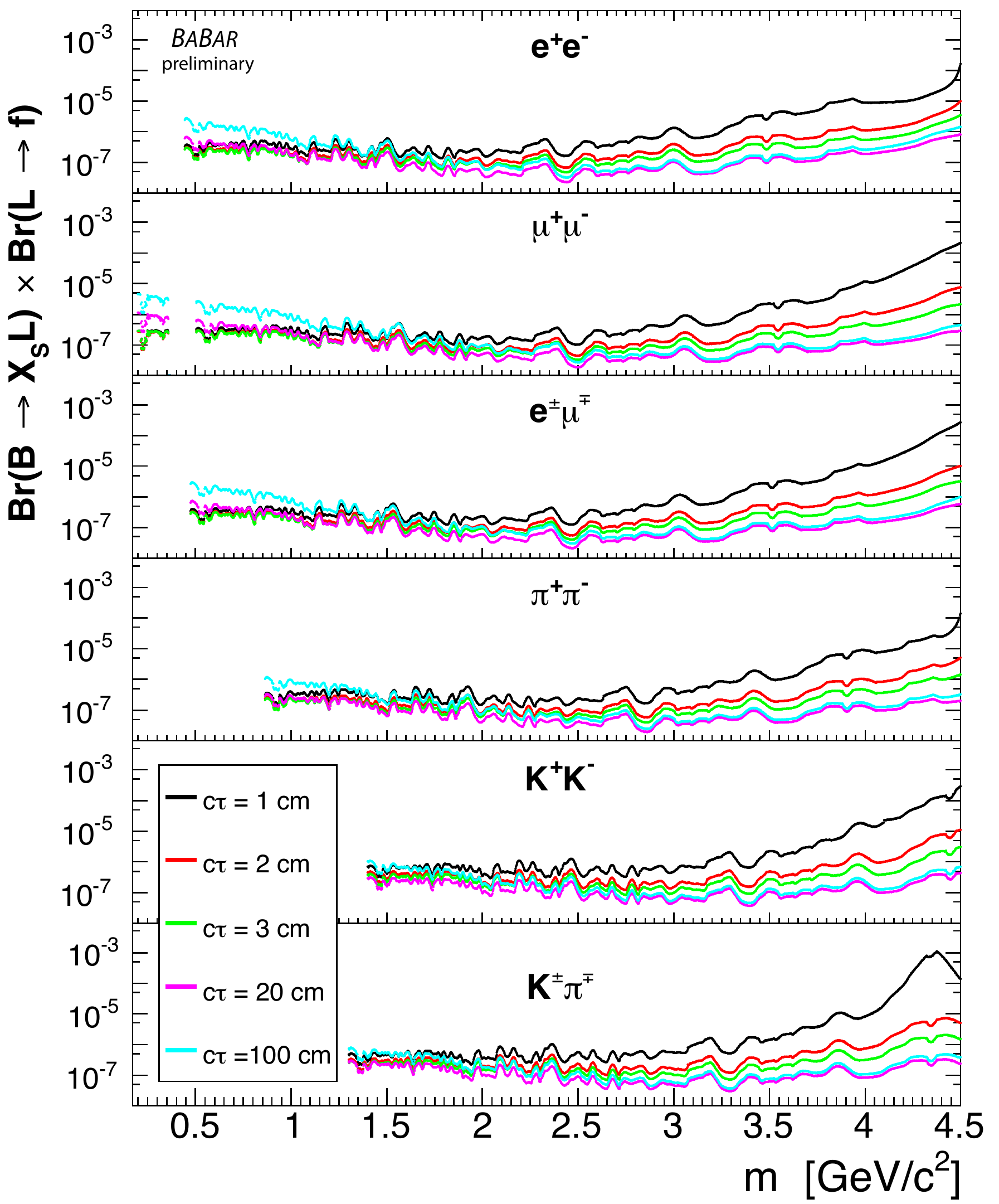}
\caption{Upper limits on the cross section $\sigma_f$ as a function of L mass for each final state  (left) for $\Upsilon(4S$) data (lower red curves) and for  $\Upsilon(3S+\Upsilon(2S)$ data (upper blue curves) and on the product branching fraction ${\cal B}_{Lf}$ (right) for different decay lengths.   }
 \label{fig:long-lived}
\end{figure}

For each scan fit, we determine the systematic error on  the signal yield. The dominant systematic error comes from the background PDF. Other
contributions are due to the mass resolution, luminosity, particle identification and the number of MC signal events. After adding all contributions in quadrature, we include the total systematic error $(\sigma_{sys})$ in the fit by convolving the likelihood function with a Gaussian function with width $\sigma_{sys}$. We determine uniform-prior Bayesian upper limits at $90\%$ CL on $\sigma_f$ as a function of $m$ for each final state. Figure~\ref{fig:long-lived} (left) shows the results on $\sigma_f$ separately for the $\Upsilon(4S)$ and $\Upsilon(3S)+\Upsilon(2S)$ data. 
Using the efficiency tables~\cite{babar15}, these  limits can be applied to any production model. For the model-dependent interpretation, we set Bayesian upper limits at $90\%$ CL  on the product of branching fractions ${\cal B}_{Lf}$. Figure~\ref{fig:long-lived} (right) shows the results on ${\cal B}_{Lf}$ as a function of $m$ for different values of $ c\tau$. These limits  exclude a significant region of the parameter space of the inflaton model~\cite{Bezrukov}.

\section{Search for New $\pi^0$-like Particles}

\begin{figure}[h]
\centering
\vskip -0.4cm
\includegraphics[width=90mm]{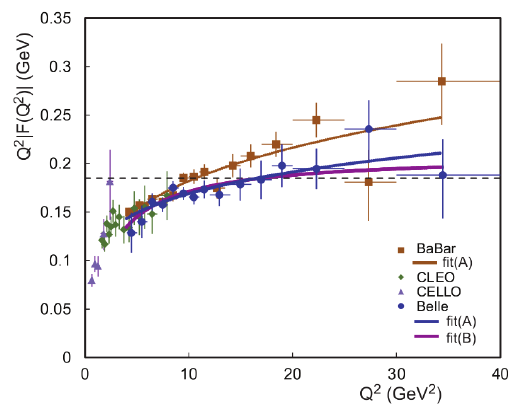}
\caption{Measurements of the $\pi^0$ form factor as a function of $Q^2$.}
 \label{fig:BL-limit}
\end{figure}

The measurement of the $\pi^0$ form factor in two-photon collisions by the \babar\ Collaboration~\cite{babar09a} has raised various discussions~\cite{Bakulev, Dorokhov, Lucha, Noguera}. At sufficiently high squared momentum transfer $Q^2$, the pion form factor should approach the Brodsky-Lepage limit of $ \sqrt{2} f_\pi/Q^2  \simeq 185~ \rm MeV/$$Q^2$~\cite{Brodski}. At $Q^2 > 15~ \rm GeV^2/c^2$, the data is expected to be described well  by perturbative QCD. Our data, however, show no sign of convergence towards the Brodsky-Lepage limit as Fig.~\ref{fig:BL-limit} indicates. Though the Belle data~\cite{belle12} show a better agreement with the perturbative prediction, they are consistent with our results. So, maybe a new particle $\phi$ exists with mass close to that of the $\pi^0$  decaying to $\gamma \gamma$  and thus causing the non-convergence of the $\pi^0$ form factor.  The new particle $\phi$ may be a scalar ($\phi_S$),  pseudoscalar ($\phi_P$) or a so-called hard-core pion ($\pi^0_{HC}$)~\cite{McKeen}.

\begin{figure}[h]
\centering
\vskip -0.2cm
\includegraphics[width=50mm]{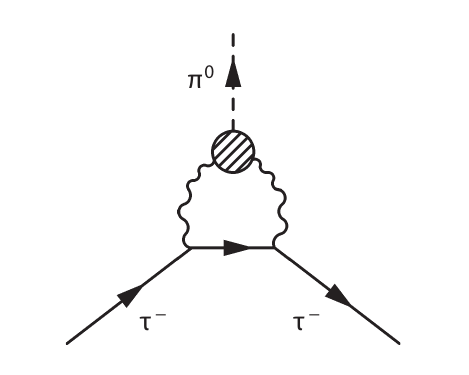}
\includegraphics[width=50mm]{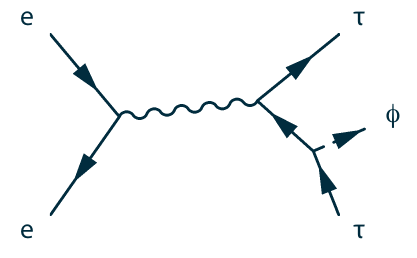}
\caption{Leading-order SM diagram for $\tau^+ \tau^-$ annihilation into a $\pi^0$-like particle (left) and for the radiation of a $\phi$ from a $\tau$ lepton in $e^+ e^- \ra \tau^+ \tau^-$  (right).}
 \label{fig:pi0-like}
\end{figure}

We focus on a search for $\pi^0$-like particles that are produced in association with a $\tau^+ \tau^-$ pair~\cite{babar15a}. 
Figure~\ref{fig:pi0-like} shows lowest-order Feynman diagrams for the production of $\pi^0$-like particles in association with  $\tau^+ \tau^-$.   For $Q^2 >8~\rm GeV^2/c^2$, predicted cross sections for $e^+ e^- \ra \tau^+ \tau^- \phi $ are large: $\sigma_{\pi^0_{HC}} =0.25$~pb, $\sigma_{\phi_P} =2.5$~pb, and $\sigma_{\phi_S} =68$~pb. Thus, we expect large event yields in our data set: $N_{\pi^0_{HC}} =1.2 \times 10^5, N_{\phi_P}= 1.2 \times 10^6$ and $N_{\phi_S}= 3.2 \times 10^7$, respectively.

\begin{figure}[h]
\begin{minipage}{19pc}
\includegraphics[width=86mm]{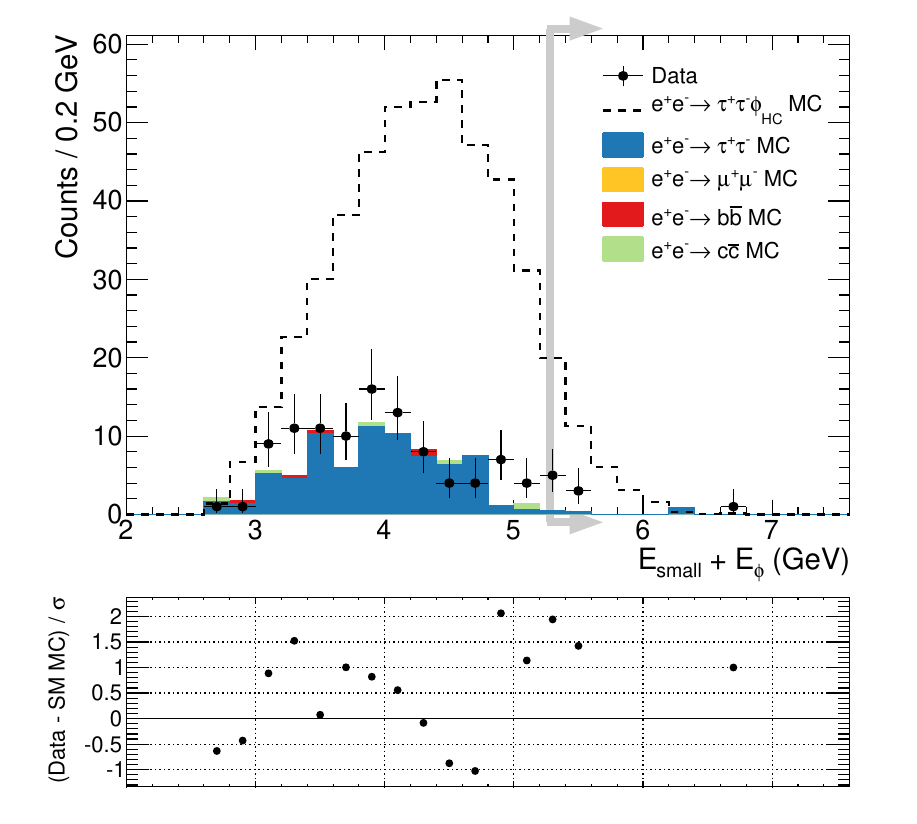}
\caption{\label{fig:Emin}Top: the  $E_{small} + E_{\phi}$ distribution for events with $100 < m_{\gamma \gamma} < 160$~\mevc2\ in data (points with error bars) and simulations for $e^+ e^- \ra \tau^+ \tau^-, ~\mu^+ \mu^-, ~b \bar b, ~c \bar c$~(colored histograms). The dashed histogram shows the expectation for $\pi^0_{HC}$ production at $\sigma= 0.254$~pb. Bottom:  pull distribution.}
\end{minipage}\hspace{2pc}%
\begin{minipage}{17pc}
\includegraphics[width=78mm]{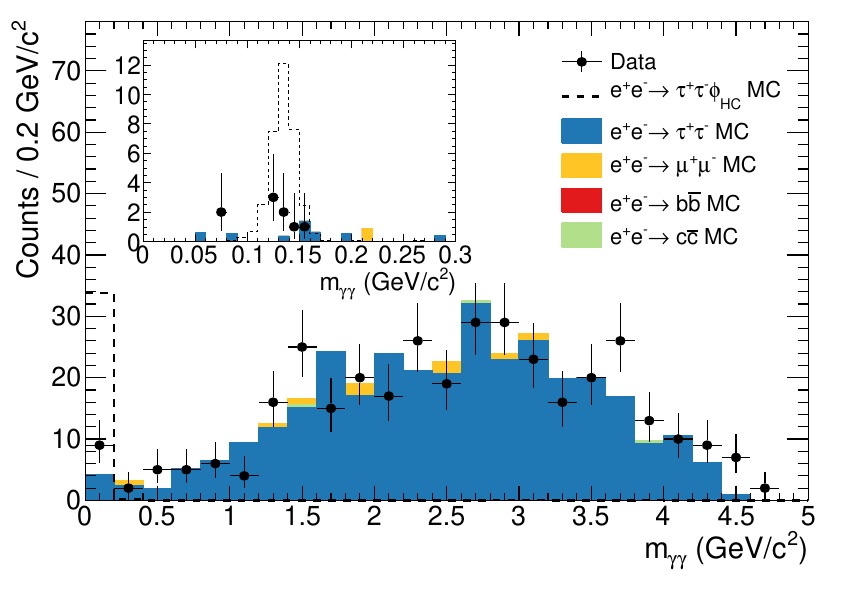}
\caption{\label{fig:mug}The $m_{\gamma \gamma} $ invariant-mass distribution (right) for data (solid points with error bars), expected signal (dashed histogram) and simulated $e^+ e^- \ra \tau^+ \tau^-, ~\mu^+ \mu^-, ~b \bar b, ~c \bar c$ background spectra (colored histograms). The inset shows the low-mass region in bins of 10~\mevc2.}
\end{minipage} 
\end{figure}

Using the full \babar\  $\Upsilon(4S)$ data set of  $468~\rm fb^{-1}~( 4.3 \times 10^8 \tau$ pairs), we select $\tau^+ \tau^-$ events in the decays 
$\tau^+ \ra \e^+ \bar \nu_e \nu_\tau$ and $\tau^- \ra \mu^-  \nu_\mu \bar \nu_\tau$\footnote{charge conjugation is implied unless stated otherwise.} 
decays requiring $p_T > 0.3~\rm GeV/c$ for each lepton. In addition, we require exactly one $\pi^0 \ra \gamma \gamma$ with energy
$  2.2 < E_{\pi^0} < 4.7$~\gev\ in the laboratory frame.  After excluding the energy of the photons from the $\pi^0$ decay, we require the remaining energy $E_{extra}$ in the electromagnetic calorimeter to be less than 0.3~\gev. We reduce background from radiative Bhabhas by imposing a minimum photon energy of $E_\gamma >0.25$~\gev\ and an opening angle between  $\e^\pm$ and $\gamma$ of $30^\circ < \theta(e, \gamma) < 150^\circ$.
We further reduce background from semi-leptonic and hadronic  $\pi^\pm \pi^0$ decays by requiring $E_{small} + E_{\phi} > 0.5 E_{CM}$ and $ m_{\pi^\pm \pi^0} > m_\tau$ where $E_{small}$ is the energy of the lower-energy track and $E_{CM}$ is the CM energy. 
Figure~\ref{fig:Emin}  shows the $E_{small} + E_{\phi}$ distribution before the requirement of $E_{small} > 0.5 E_{CM}$, which is consistent with the expected  $e^+e^- \ra \tau^+ \tau^-$ energy spectrum. We simulate background modes $e^+ e^- \ra B \bar B$ with EvtGen~\cite{Lange}, $e^+ e^- \ra q \bar q$ continuum with JETSET~\cite{JETSET}, $e^+ e^- \ra e^+ e^-$ with BHWIDE~\cite{BHWIDE},   $e^+ e^- \ra \mu^+ \mu^-, \tau^+ \tau^-$ with KK2F~\cite{KK2F} where we use the TAUOLA library~\cite{TAUOLA} to generate $\tau$ decays. We model radiative corrections with PHOTOS~\cite{PHOTOS} and the simulation of the detector response with GEANT4~\cite{GEANT4}.

\begin{figure}[h]
\begin{minipage}{17pc}
\includegraphics[width=79mm]{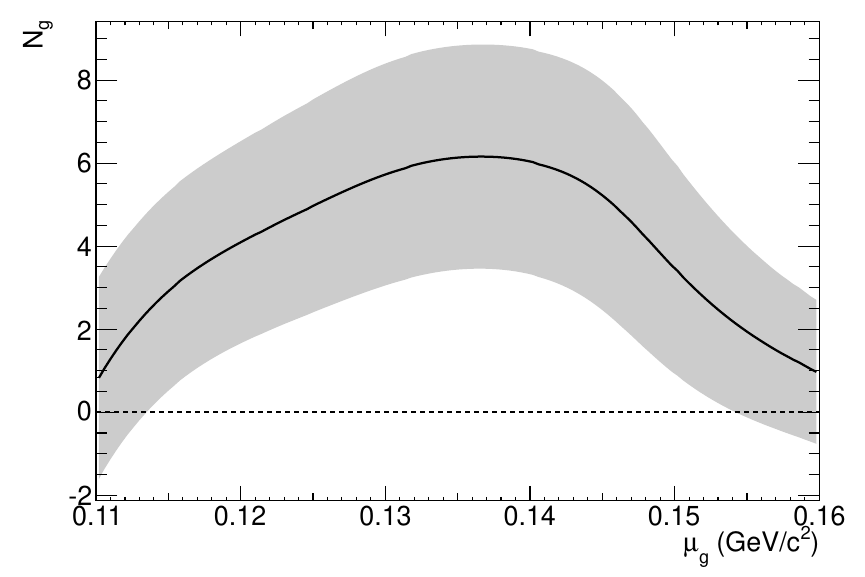}
\caption{\label{fig:pi0-limit}Number of events $N_g$  as a function of the $\phi$ mass hypothesis $\mu_g$}
\end{minipage}\hspace{2pc}%
\begin{minipage}{19pc}
\includegraphics[width=79mm]{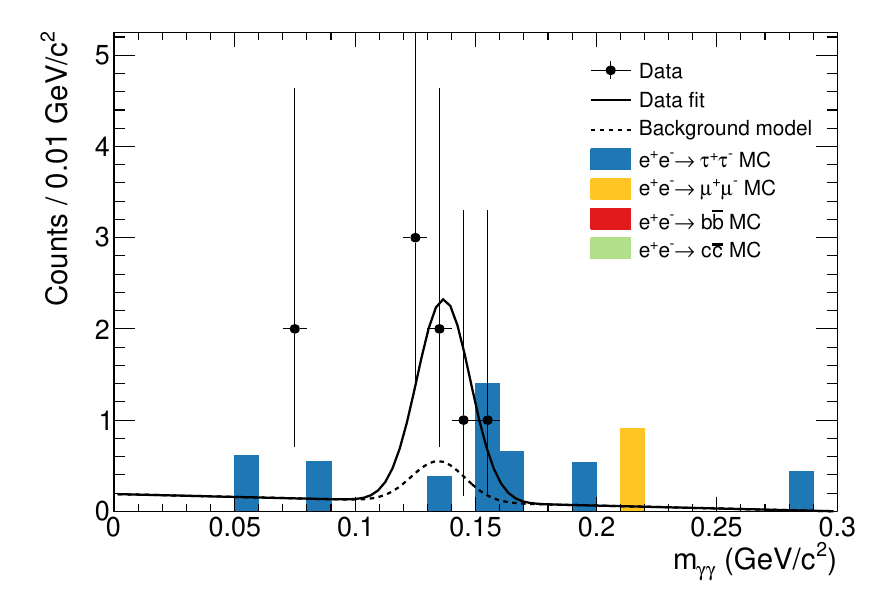}
\caption{\label{fig:pi0-fit}Measured $m_{\gamma \gamma}$ spectrum (points with error bars) with fit overlaid (solid line), peaking background contribution (dotted line) and expected backgrounds from $e^+ e^- \ra \tau^+ \tau^-, ~\mu^+ \mu^-, ~b \bar b, ~c \bar c$  (colored histograms).}
\end{minipage} 
\end{figure}

Figure~\ref{fig:mug} shows the entire $m_{\gamma \gamma}$ invariant-mass spectrum after applying all selection criteria. For comparison, we also show a simulated signal and the simulated spectra for $e^+ e^- \ra \tau^+ \tau^-,~ \mu^+ \mu^-,~ b \bar b$ and $ c \bar c$ backgrounds. We fit the $m_{\gamma \gamma}$ invariant-mass distribution in the region 50 $< m_{\gamma \gamma} <$ 300~\mevc2\ with a Gaussian signal and a linear background: $N(m_{\gamma \gamma}) = N_g~ G(\mu_g, \sigma_g) +  N_{lin} ~(1+a_{lin}~ m_{\gamma \gamma})$. We use extended unbinned maximum log-likelihood fits to extract  $N_{lin}, N_g$ and $a_{lin}$. The width  $\sigma_g$ of the Gaussian function is fixed to 
11.1~\mevc2\ determined from control samples. The mean $\mu_g$ represents the mass of the $\phi$ particle and is fixed in each fit. 
We scan $\mu_g$ values between 110~\mevc2\ and 160~\mevc2\ in steps of 0.5~\mevc2\ and extract $N_g$ from each fit. Figure~\ref{fig:pi0-limit} shows the resulting $N_g$ distribution.  Figure~\ref{fig:pi0-fit} shows the fit for the highest yield.  We observe a raw yield of $N_g = 6.2 \pm 2.7 \pm 0.06$ events at a mass of 137~\mevc2. After subtracting $1.24\pm 0.37$ events of peaking background and correcting for a fit bias of $-0.06 \pm 0.02$ events,  we find $5.0 \pm 2.7 \pm 0.4$ signal events. We determine identical efficiencies of $(0.455 \pm 0.019)\%$ for $\pi^0_{HC}$ and $\phi_P$.  For $\phi_S$, the efficiency is reduced to $(0.09\pm 0.004)\%$. Systematic uncertainties include contributions from the MC sample size, energy scale, $\pi^0$ efficiency, energy resolution, particle identification, momentum scale and momentum resolution and amount to $4.2\%$  for $\pi^0_{HC}, \phi_P$ and $4.4\%$ for  $\phi_S$.
This leads to measured cross sections of $\sigma_{\pi^0_{HC}, \phi_P}=(38 \pm 21_{stat} \pm 3_{sys})~ \rm fb$ and $\sigma_{\phi_S} =(190\pm 100_{stat} \pm 20_{sys})~ \rm fb$, respectively. Since the result is consistent with background, we set Bayesian upper limits on the cross sections  assuming a flat prior. We obtain $\sigma_{\pi^0_{HC},\phi_P} < 73~ \rm fb$ and $\sigma_{\phi_S} < 370~ \rm fb$ at $90\%$ CL. This should be compared to predictions of $270$ -- $820~\rm fb$ and $68$ -- $ 1850~\rm fb$, respectively.

To check the compatibility of the measured production cross sections with the contribution of particle $\phi$ to the $\pi^0$ form factor, we include
this measurement as an additional term in the  $\chi^2$
of the fit to the $Q^2$ dependence of the $\pi^0$ in our data. 
The increase in $\chi^2$ follows a $\chi^2$ distribution with one degree of freedom. Thus, we can determine $p-$values for the process $e^+ e^- \ra \tau^+ \tau^- \phi$ to contribute to the $\pi^0$ form factor. We obtain $p_{\pi^0_{HC}}=5.9 \times 10^{-4}$ for $\pi^0_{HC}$, $p_{\phi_P}=8.8 \times 10^{-10}$ for $\phi_P$ and  $p_{\phi_S}=2.2 \times 10^{-9}$ for $\phi_S$. We conclude that a new particle $\phi$ with mass close to the $\pi^0$ is not a likely explanation of the  excess of the $\pi^0$ form factor seen in our data at larger $Q^2$.

\section{Conclusion and Outlook}

The $B$ factories are an excellent laboratory to search for light dark matter. We see no dark photons in the  $0.02- 10.2$~\gevc2\ mass region pushing the limit on the mixing parameter $\epsilon$ at a level of $10^{-4}$ to $10^{-3 }$ depending on the dark photon mass. We performed the first search for long-lived particles at a high-luminosity $e^+e^-$ collider and the first search in a heavy-flavor environment. We observe no long-lived particle in the $0.2  < m < 10$~\gevc2\ mass range and proper decay lengths of $ 0.5 < c\tau < 100$~cm.  In a model-independent approach, we set $90\%$ CL upper limits on the product $\sigma( e^+ e^- \ra L X) \cdot {\cal B}(L \ra f) \cdot \epsilon_f$ for different masses, $p_T$ values and  proper decay lengths. We also give model-dependent $90\%$ CL upper limits on the product branching fraction ${\cal B}(B\ra L X_s) \cdot {\cal B}(L \ra f)$. Finally, we searched for the production of $\pi^0$-like particles in association with a $\tau^+ \tau^-$ pair and set upper limits on $\sigma(e^+ e^- \ra \tau^+ \tau^- \phi)< 73$~fb at $90\%$ CL. The hypothesis that a $\pi^0$-like particle causes the non convergence of the pion form factor has a p-value of $5.9 \times 10^{-4}$ for $\pi^0_{HC}$  and thus is rather unlikely. For $\phi_P$ and $\phi_S$ production, the $p$-value is even lower. \babar\ will continue with searches for new physics.  Belle just released first limits on  $e^+ e^- \ra A^\prime h^\prime, h^\prime \ra A^\prime A^\prime$~\cite{belle15}. Since Belle's luminosity is about a factor of two larger, the limits are a factor of two lower than the \babar\ results. However, substantial improvements are expected from Belle II. At 50~$\rm ab^{-1}$, present \babar\ limits will be improved by up to two orders of magnitude depending on the final state.

\section{Acknowledgment}
This work has been supported by the Norwegian Research Council. I would like to thank the \babar\ collaboration for the opportunity to give this talk. In particular, I would like to thank Bertrand Echenard, Frank Porter and Abi Soffer  for useful comments.

\smallskip

\end{document}